\documentstyle{article}
\def\beq{\begin{eqnarray}}
\def\eeq{\end{eqnarray}}
\def\bea{\begin{eqnarray*}}
\def\eea{\end{eqnarray*}}

\def\centeron#1#2{{\setbox0=\hbox{#1}\setbox1=\hbox{#2}\ifdim
\wd1>\wd0\kern.5\wd1\kern-.5\wd0\fi
\copy0\kern-.5\wd0\kern-.5\wd1\copy1\ifdim\wd0>\wd1
\kern.5\wd0\kern-.5\wd1\fi}}
\def\ltap{\;\centeron{\raise.35ex\hbox{$<$}}{\lower.65ex\hbox{$\sim$}}\;}
\def\gtap{\;\centeron{\raise.35ex\hbox{$>$}}{\lower.65ex\hbox{$\sim$}}\;}
\def\gsim{\mathrel{\gtap}}
\def\lsim{\mathrel{\ltap}}

\def\singleandthirdspaced{\baselineskip=\normalbaselineskip\multiply
    \baselineskip by 130\divide\baselineskip by 100}

\newcommand{\newc}{\newcommand}
\newc{\qbar}{{\overline q}}
\newc{\Kahler}{K\"ahler }
\newc{\deltaGS}{\delta_{\rm GS}}

\def\lrf#1#2{ \left(\frac{#1}{#2}\right)}
\def\lrfp#1#2#3{ \left(\frac{#1}{#2} \right)^{#3}}
\newcommand{\la}{\left\langle}
\newcommand{\ra}{\right\rangle}

\begin{document}
\begin{titlepage}
\begin{flushright}
{\large hep-th/yymmnnn \\
SCIPP-10/06\\
IPMU10-0073\\
}
\end{flushright}

\vskip 1.2cm

\begin{center}

{\LARGE\bf Discrete R Symmetries and Domain Walls}

\vskip 1.4cm

{\large  Michael Dine$^{a,b}$, Fuminobu Takahashi$^c$ and Tsutomu~T. Yanagida$^{c,d}$}
\\
\vskip 0.4cm
{\it $^a$Santa Cruz Institute for Particle Physics and
\\ Department of Physics, University of California,
     Santa Cruz CA 95064  } \\
 {\it $^b$Department of Physics, Stanford University
Stanford, CA 94305-4060, USA}\\
{\it $^c$Institute for the Physics and Mathematics of the Universe, \\University of Tokyo,
Chiba 277-8583, Japan  }\\
{\it $^d$ Department of Physics, University of Tokyo,
     Tokyo 113-0033, Japan\\
}

\vskip 4pt

\vskip 1.5cm

\begin{abstract}
Discrete $R$ symmetries are interesting from a variety of points of
view.  They raise the specter, however, of domain walls, which may be
cosmologically problematic.  In this note, we describe some of the
issues.  In many schemes for supersymmetry breaking, as we explain,
satisfying familiar constraints such as suppression of gravitino production,
insures that the domain walls are readily inflated away.  However, in
others, they form after inflation.  In these cases, it is necessary
that they annihilate.  We discuss possible breaking mechanisms for the
discrete symmetries, and the constraints they must satisfy so that the
walls annihilate effectively.
\end{abstract}

\end{center}

\vskip 1.0 cm

\end{titlepage}
\setcounter{footnote}{0} \setcounter{page}{2}
\setcounter{section}{0} \setcounter{subsection}{0}
\setcounter{subsubsection}{0}

\singleandthirdspaced

\section{Domain Walls and Discrete $R$ Symmetries}
\label{sec:1}

If supersymmetry (SUSY) plays a role in low energy physics, discrete
symmetries seem likely to be an important component.  $R$ parity is
one important example.  But more generally, there are a number of
reasons to think that {\it discrete R symmetries} should play a
significant role.  For example, approximate, continuous $R$ symmetries
seem an essential component of dynamical supersymmetry breaking, and
one way these might arise is as an accidental consequence of discrete
$R$ symmetries.  Such symmetries might play a role in suppressing
dimension five operators, and accounting for the scale of
supersymmetry breaking.  Note here that when we speak of $R$
symmetries, we are excluding simple $Z_2$ symmetries (like $R$ parity)
which at most rotate the phase of the supercharges by $\pi$; such
symmetries can always be redefined by a $2\pi$ rotation so as to leave
the supercharges alone.

Any discrete $R$ symmetry, however, must be spontaneously broken.
This is because a non-vanishing superpotential is required in the
effective action at scales of order the supersymmetry breaking scale
in order to account for the small value of the cosmological constant,
and this breaking must be substantial.  Domain walls are then
inevitable.
 These domain walls are problematic cosmologically~\cite{Zeldovich:1974uw,Kibble:1976sj}, and
either must be inflated away, or, if the symmetry is not exact, must
rapidly annihilate~\cite{Vilenkin:1981zs}.  In this note, we consider these issues carefully.
We first survey the scales of $R$ symmetry breaking in different
schemes for supersymmetry breaking.  We will see instances where other
constraints, such as overproduction of gravitinos~\cite{Moroi:1993mb,de Gouvea:1997tn,Kawasaki:2004qu,Jedamzik:2006xz}, insure that the
discrete symmetry is broken at a scale well above the reheating
temperature; then domain walls have been inflated away provided the
scale of inflation is not much below the GUT scale.  However, in
others, the problem is serious; the reheating temperature can be high,
restoring the symmetry, or, if not, solving the problem with inflation
requires inflation at a rather low scale ($10^{13}$ GeV or smaller.)
So it is necessary to consider the possibility that the walls
annihilate, i.e. that the discrete symmetries are not exact.  We note
that, in string theory, small, explicit breaking of discrete $R$
symmetries seems common, and we determine conditions under which the
domain walls annihilate sufficiently rapidly.

As we will see, in the case of intermediate scale supersymmetry
breaking ($m_{3/2} \sim {\rm TeV}$), the likely scales of $R$
breaking range from $10^{13}$\,GeV to $M_p$.  At the upper end, domain
walls are even more catastrophic than conventionally assumed.  These
walls are parameterically far more problematic than the usual moduli
problem.  At the lower end, assuming that the usual gravitino problem
of such theories is solved, the domain wall problem is readily solved
as well.  In low scale supersymmetry breaking, the fate of domain
walls is a function of the supersymmetry breaking scale.  For a broad
range of $m_{3/2}$, gravitino overabundance is a severe
problem~\cite{Moroi:1993mb,de Gouvea:1997tn,Kawasaki:2004qu,Jedamzik:2006xz}, and solving that problem requires a
relatively low reheating scale, well below the scale of discrete $R$
breaking.  Avoiding domain walls then constrains the value of the
Hubble parameter during inflation to be below the $R$ breaking scale,
and one finds again that the domain wall problem is solved without
terribly low scale inflation.  However, for very low scale gauge
mediation, the reheating temperature is not significantly constrained,
and the domain wall problem, as we will see, is severe unless the scale of
inflation is quite low.

The rest of this paper is organized as follows.  In the next section,
we discuss the scales of $R$ symmetry breaking in different scenarios
for supersymmetry breaking, and explain under what circumstances
domain walls are problematic.  In section \ref{annihilation}, we
consider the problem of domain wall annihilation.  We explain that in
string theory, it is (in some sense) common for discrete symmetries to
be explicitly broken by a small amount, and determine the conditions
under which annihilation is sufficiently rapid to avoid cosmological
problems.  In section \ref{sec:4} we briefly discuss the retrofitted models.
In section \ref{gravitywaves}, we consider the possibility
that observable gravitational wave signals might emerge from domain
wall collisions.

\section{Scales of R Symmetry Breaking and the Problem of Domain Walls}
\label{rscales}

The universe may have undergone multiple inflationary periods.
Assuming that the last inflation has continued for sufficiently long,
domain walls are formed in our observable universe if the Hubble
parameter during the last inflation $H_I$ exceeds the $R$-breaking scale $\Lambda$,
or alternatively if the highest temperature of the $R$-breaking sector
after inflation, $T_H$, exceeds $\Lambda$~\footnote{
The highest temperature $T_H$ is related to the reheating temperature and
the inflation scale as $T_H \simeq (T_R^2 H_I M_p)^{1/4}$.
}.  Thus, the walls are formed
if the following inequality is met;
\beq
{\rm max}[H_I,T_H] \gsim \Lambda.
\label{inf_scale}
\eeq
 We will see in this section that avoiding domain wall formation, in
 some cases, is compatible with a relatively large scale of inflation,
 but the upper bound on the scale can also be quite low, possibly
 lower than $10^{9}$ GeV. Various possibilities for achieving
 inflation at these scales within supersymmetric models have been
 discussed in the  literature~\cite{Kumekawa:1994gx,Izawa:1996dv,Asaka:1999jb,Copeland:1994vg}.

 Without a detailed model, the only information one has about the
 scale of discrete $R$ breaking comes from the relation:
 \beq \vert
 \langle W \rangle \vert = {1 \over \sqrt{3}} \vert F \vert M_p,
\eeq
where $F$ is the gravitino decay constant.
$W$ itself is
 an order parameter of $R$ symmetry breaking, but potentially there
 are a variety of order parameters for $R$ symmetry as
 well~\cite{dinekehayias}.  Roughly speaking, we are interested in two
 scales.  The first we will call $M_r$, the scale of $R$ symmetry
 breaking. This scale corresponds to the masses of particles which
 gain mass as a consequence of $R$ symmetry breaking.  The second is
 $m_r$, loosely speaking the mass of the particles whose dynamics is
 responsible for $R$ symmetry breaking.  If we call
\beq \Lambda=
 \vert W \vert^{1/3} \approx 8 \times 10^{12} {\rm\, GeV}
 \lrfp{m_{3/2}}{100{\rm GeV}}{1/3},
\eeq
$\Lambda$ is not necessarily equal to either $M_r$ or $m_r$.  Within
various standard pictures for supersymmetry breaking, we can enumerate
possible values of $M_r$ and $m_r$:
\begin{enumerate}
\item  Intermediate scale supersymmetry breaking (``supergravity breaking"), $R$ symmetry broken in hidden sector:
Here, we can distinguish two cases.   Given
that we are supposing an underlying discrete $R$ symmetry, this symmetry might be carried by the hidden sector
fields.  In this case, we would expect $M_r = M_p$, $m_r = m_{3/2}$.
\item  Intermediate scale supersymmetry breaking, $R$ symmetry broken by additional interactions (retrofitting):  The discrete $R$ symmetry might be broken by some other dynamics, as in retrofitted models~\cite{retrofitting},  Then we might have $M_r = \Lambda = m_r$.
\item  Low scale supersymmetry breaking (gauge mediation):  In this case, the $R$ symmetry
is inevitably broken by some additional dynamics at a scale much larger than that of supersymmetry breaking~\cite{Yanagida:1997yf, dinekehayias,wtheorem}.  Within
the framework of ``retrofitted" models, one might expect that $M_r = \Lambda = m_r$.  The cosmology of the
resulting domain walls is then quite sensitive to $F$ (or $m_{3/2}$), the supersymmetry breaking order parameter.
\end{enumerate}

In the following subsections, we consider each of these cases in turn.

\subsection{Intermediate Scale Supersymmetry Breaking}

In most scenarios for supersymmetry breaking at an intermediate scale,
supersymmetry is broken in a hidden sector.  The
longitudinal mode of the gravitino is assumed to arise from a chiral
field, whose scalar component is a pseudomodulus with mass of order
$m_{3/2}$.  For definiteness, we will describe a situation where there
is one such field, $Z$.  If $Z$ transforms under the discrete
symmetry, then the superpotential has the form, for small $Z$,
\beq W
= m_{3/2} M_p^2\left [ \left ({Z \over M_p} \right)^a \left (1 + c_N
  \left ({Z \over M_p} \right)^N + \dots \right) \right ],
\eeq
where
we have assumed that the discrete symmetry is $Z_N$.  We allow a
general Kahler potential consistent with the symmetry.  Examining this
expression, one sees that in order that the cosmological constant be
small, it is necessary that \beq \langle Z \rangle \sim M_p.  \eeq So
the breaking of the discrete symmetry is necessarily of order $M_p$.
So, indeed, $M_r \sim M_p$, $m_r \sim m_{3/2}$.

The domain wall tension in theories of this type is of order $m_{3/2}
M_p^2$.  This can be seen by simple scaling arguments.
Cosmologically, this is highly problematic.  Models with pseudomoduli
with masses of order $m_{3/2}$ already have severe cosmological
problems.  When $H \sim m_{3/2}$, these moduli simultaneously begin to
oscillate and also dominate the energy density of the universe.  There
are two possible behaviors for the domain walls in such systems:
\begin{enumerate}
\item During inflation, the field $Z$ might be driven to a point in
  field space, far away from its final stationary point, but at which
  the discrete symmetry is already broken; the domain walls which
  exist at this stage will be inflated away.  There is no need for
  further domain walls to form in the postinflationary dynamics of
  $Z$.
\item During inflation, the field $Z$ might sit at a point where the
  discrete symmetry is unbroken.  Domain walls, then, {\it will} form
  after inflation ends.  As the field settles into its minimum, with
  $H \approx m_{3/2}$, one might expect there to be of order one
  domain wall per horizon.  The energy stored in this wall would be of
  order $m_{3/2}^{-1} M_p^2$, corresponding to an energy density of
  order $m_{3/2}^2 M_p^2$, parameterically as large as the energy
  stored in the field!  In other words, at this stage, the domain
  walls would dominate the energy density.  Clearly this is
  cosmologically unacceptable, unless, somehow, the domain walls can
  rapidly annihilate.  We will discuss this possibility in section
  \ref{annihilation}.
\end{enumerate}

An alternative is that the breaking of the $R$ symmetry arises in a
different sector, as in retrofitted models.  The parameter, $m_{3/2}
M_p^2$ in the superpotential might arise through a coupling such as
\beq
\int d^2 \theta {Z W_\alpha^2 \over M_p},
\eeq
where $W_\alpha$ is the field strength of a new gauge group, with
scale $\Lambda$.  In this case,
\beq M_r = m_r = \Lambda,
\eeq
and $\Lambda = (m_{3/2} M_p^3)^{1/4} \approx 10^{14}$\,GeV for $m_{3/2} = 100$\,GeV.  This scale is quite large~\footnote{
If the R breaking occurs in the squark condensation instead of gaugino condensation,
we can raise the R-breaking scale $\Lambda$ up to any very large value. For instance consider $W=({\bar Q }Q)^n/M_p^{2n-3}$ with  $\la {\bar Q}Q \ra=\Lambda^2$ and the R-charge of $({\bar Q}Q)^n$ equal to $2$ modulo $N$. Then we get $\Lambda ^{2n} = m_{3/2}M_{p}^{2n-1}$.
Taking $n$ sufficiently large we can easily make $\Lambda > H_{I}$.}; the
domain wall tension would be of order $\Lambda^3$.  But what is most
important for the question of domain walls is the value of $H$ during
inflation.  Necessarily, in models such as this, the reheat
temperature is less than $10^9$\,GeV~\cite{Kawasaki:2004qu,Jedamzik:2006xz}.  So the $R$-symmetry breaking
phase transition must complete before reheating.  The question of
whether the domain walls inflate away is the question of whether,
during inflation, $H \equiv H_I$ is greater than $\Lambda$.  If it is,
then the transition may well not be completed during inflation, and
the possibility of dangerous domain walls exists.  If $H_I < \Lambda$,
then the domain walls will be inflated away.  The latter condition
corresponds to an energy scale at inflation, $E_I  \simeq \sqrt{H_I M_p} \approx 10^{16}$\,GeV,
i.e. as long as the scale of inflation is below $10^{16}$\,GeV, the
domain walls will inflate away.\footnote{The WMAP 7-year data sets an
  upper bound on the inflation scale, $H_I \lsim 1.6 \times
  10^{14}\, {\rm GeV}$~\cite{Larson:2010gs}.  High-scale inflation
  models such as chaotic inflation~\cite{Linde:1983gd,Kawasaki:2000yn} satisfying $H_I > \Lambda$
  may produce tensor modes which can be measured
   by the future CMB observations.  }

There is an interesting possibility that the domain walls induce a topological inflation.
That is, our universe is contained in a domain wall during the inflation and there will be no walls
in the observable universe after inflation. The model is given as follows.
We suppose that the discrete $R$ symmetry is $Z_{4R}$ and $X$ and $Z$ carry the $R$ charge $2$.
The superpotential is
\beq
W\;=\; v^2 X \left( 1- g\lrfp{Z}{M_P}{2} \right) + \sqrt{g} m_{3/2}M_p Z,
\eeq
and $Z$ gets a vev, $\la Z \ra \simeq M_p\sqrt{g}$~\cite{Izawa:1998rh}. The domain walls are generated in association with the $R$-symmetry breaking.
We see that the topological inflation occurs in a domain wall if the coupling $g$ is $O(1)$ and that
the observed density perturbation is explained  for $v=10^{13}-10^{15}$ GeV.
The difference from the original model of Ref.~\cite{Izawa:1998rh} is that
the large superpotential is generated by the Planck scale vev of $Z$.
The linear term of $Z$ may be originated from the dynamics which breaks supersymmetry.

Finally, there is the possibility that there are no pseudomoduli
fields in the hidden sector.  This case could arise if supersymmetry
is broken without pseudomoduli, as in the 3-2
model~\cite{Affleck:1984xz}.  However, in such a situation, there
still must some additional dynamics responsible for the large $W$
needed to cancel (the bulk of) the cosmological constant.

To summarize, in the case of intermediate scale supersymmetry
breaking, the first question to ask is whether the hidden sector
fields are the source of $R$ symmetry breaking.  If they are, the next
question is whether the symmetry is already broken during inflation or
not.  If not, one needs to explore the possibility of domain wall
annihilation.  In the event that some other dynamics are responsible
for breaking the $R$ symmetry, there need not be a domain wall
problem.

\subsection{Low scale supersymmetry breaking}

One of the traditional objections to gauge
mediation~\cite{banksunpublished} is that it is hard to understand how
one generates the large superpotential necessary to cancel the
cosmological constant.  If there is a discrete $R$ symmetry, some
additional dynamics, such as gaugino condensation, is needed.  As
noted in refs. \cite{dinekehayias,retrofitting}, this can be quite
natural, if the role of the additional dynamics is to generate the
scales in an O'Raifeartaigh model.  In such a case, one again has
\beq
M_r = m_r =\Lambda.
\eeq
(This is the case even if the model is not retrofitted, and one simply
has gaugino condensation of something similar to generate $\langle W
\rangle$~\cite{Yanagida:1997yf}.)  So we must ask how large is $\Lambda$ for a given scale of
supersymmetry breaking, and then determine for what value of $E_I$ the
resulting domain walls are inflated away.

Let's start with extremes.  Consider a relatively high scale for
supersymmetry breaking, $m_{3/2} = 100$\,MeV.  In this case,  $\Lambda =
(m_{3/2} M_p^2)^{1/3} \approx 10^{12}$\,GeV.  In such a case, the gravitino overabundance already
requires that the $T_R < 10^6$\,GeV.  So, again, the transition to the
broken symmetry phase must occur before reheating.  Now requiring $m_r
> H_I$ leads to $E_I < 10^{16}$\,GeV.  So rather high scales of
inflation are permitted.  At the other extreme, suppose $m_{3/2}
\approx 10$\,eV.  In this case, $\Lambda \approx 10^{9}$ GeV.  In
this case, there is no significant constraint on the reheating
temperature, so one can be concerned that $T_R > \Lambda$.  Even if
not, the condition on the scale of inflation is now $E_I < 10^{14}$
GeV.

\section{Explicit R Breaking and the Fate of Domain Walls}
\label{annihilation}

In cases where inflation occurs before the formation of domain walls, their disappearance might be explained by explicit breaking of the
symmetry~\cite{Vilenkin:1981zs,wilczeketal}. It might seem troubling to postulate a symmetry, and then invoke small, explicit breaking, but this phenomenon is rather common in string theories, where discrete symmetries (and discrete $R$ symmetries in particular) are often
anomalous~\cite{banksdineanomalies}.\footnote{ If it is not anomalous, one may consider a gauged $Z_{N R}$
  symmetry. In this case the domain walls are not formed, or, even if they are formed,
  they will disappear in the end.  However it is not easy to have such a non-anomalous $Z_{NR}$
  symmetry~\cite{Kurosawa:2001iq}.  }  In these cases, there is
typically an axion-like field which transforms non-linearly under the
discrete symmetry, whose couplings cancel the would-be anomaly.  Thus
there is an exact symmetry, which is spontaneously broken at a high
scale, and an approximate symmetry at low energies.  The effects of
the breaking at low energy are exponentially small if appropriate
couplings are small.  As the scale of spontaneous breaking of the
exact symmetry could readily be the Planck scale or the the GUT scale,
one does not need to worry about domain walls arising from the
breaking of the underlying, exact symmetry.  The breaking of the
approximate discrete symmetry at low energies has the potential to
produce problematic domain walls.  On the other hand, the
exponentially small effects associated with the anomaly will generate
a small splitting in the energies of the different
domains.  The question, then, is how large is the splitting and how quickly the domain walls
annihilate.

An essentially equivalent phenomenon can occur if, for example, there
are two gaugino condensates, one breaking a symmetry $Z_N$ at the
scale $\Lambda$, and another a symmetry $Z_{N^\prime}$ at a lower
scale $\Lambda^\prime (< \Lambda)$.  From a more microscopic point of
view, these symmetries are incompatible (all gauginos must transform
under any $R$ symmetry), and should again be thought of as
anomalous. One can view the gaugino condensate of the higher scale
theory as accounting for the size of of $\la W \ra$, while the other
lower scale condensate generates the mass splitting.

In either case, we can represent the effects of the explicit
$R$-symmetry breaking through a constant, $w_c$, in the
superpotential, and consider the effects of the spontaneous breaking
to go as $W_0= \Lambda^3 \alpha^k$, where $\Lambda$ is some dynamical
scale, $\alpha$ is a suitable root of unity, and $\Lambda^3 > w_c$.
Domain walls are produced at the phase transition associated with the
scale $\Lambda$, and they will annihilate as a result of the explicit
breaking $w_c$. Calling $w_c = a\, m_{3/2} M_p^2$, and $\Lambda^3 = b\, m_{3/2} M_p^2$, we have $a + b =
{1}$. (Here and in what follows we drop $\alpha^k$.)  We take $a \ll b \simeq 1$.

The splitting between states then behaves as
\beq
\epsilon = a\,b\,m_{3/2}^2 M_p^2.
\label{split}
\eeq
Now the value of $H$ when the walls collide can be estimated as
follows.  Calling $x$ a wall coordinate, and adopting the notation of
Vilenkin~\cite{Vilenkin:1981zs} in which $\sigma$ is the wall tension, we
have
\beq
\sigma \ddot x = \epsilon
\eeq
and $\sigma \approx b
\,m_{3/2} M_p^2$.
\beq x \approx {1 \over 2} {\epsilon \over \sigma}
t^2
\eeq
giving, for the condition $x \approx H^{-1}$,
\beq H \approx a m_{3/2}.
\label{collapse}
\eeq
The requirement that, at this time, the Schwarschild radius $r_s$
associated with the domain wall tension in a horizon be smaller than
the horizon gives the condition:
\beq
H^{-1} \gg r_s
\label{condition1}
\eeq
or
\beq
H \gg b m_{3/2}
\eeq
from which we have
\beq
b \ll a \simeq 1.
\eeq
This contradicts our original assumption.  A picture in which domain
walls form with characteristic scale $\Lambda$, and annihilate due to
some smaller, explicit $R$ breaking, is not viable.  At best, walls
connected with the lower scale of symmetry breaking can annihilate as
a result of the splitting between the domains generated by the higher
scale dynamics.

As an example, consider the case of two gaugino condensates, one
associated with scale $\Lambda$, one with $\Lambda^\prime$.  if $w_c$
arises from gaugino condensation at a scale $\Lambda^\prime$ (i.e.,
$w_c \sim \Lambda^{\prime 3}$), there are two kinds of walls. The
precise value of the mass splitting then depends on the combination of
the two vacua. In the above example with $Z_N$ and $Z_{N^\prime}$, the
typical magnitude of the bias is still given by (\ref{split}), and
there is a unique vacuum if $N$ and $N^\prime$ are coprime with
respect to each other.  The lower scale domain walls annihilate when
$H \sim b m_{3/2}$, which is earlier than
(\ref{collapse})\footnote{Even if some of the lower scale walls do
  not disappear at this time, they do not affect the subsequent
  evolution of the higher scale walls.}.  This is because the walls
with a smaller tension annihilate earlier for the same splitting. The
subsequent dynamics of the higher scale walls are the same as
described above.

\section{Domain Walls in Gauge Mediation}
\label{sec:4}

Gauge mediated models with retrofitting as the origin of supersymmetry
breaking raise similar issues to those in the gravity mediated case
(with retrofitting).  Again, one seems to require that the explicit
breaking be rather large.

In \cite{dinekehayias}, it was argued that a natural way in which to
understand dynamical supersymmetry breaking was to suppose that
gaugino condensation (in a generalized sense discussed there) at a
scale of order $(F M_p)^{1/3}$ generated a dynamical scale responsible
for supersymmetry breaking (with Goldstino decay constant $F$).  This
naturally correlated the scale of supersymmetry breaking and the need
for a large $\langle W \rangle$ needed to obtain a small cosmological
constant~\cite{Yanagida:1997yf}.  But we see that if the domain walls
associated with this condensation are to be eliminated through an
explicit breaking of the symmetry, the scale of these ``retrofitting"
interactions must be lower than $(M_p F)^{1/3}$.

It is worth recalling the basic ideas of retrofitted models.  Here one
starts with, say, a conventional O'Raifeartaigh model,
\beq
X (A^2 - \mu^2) + mAY
\eeq
and accounting for the scales $\mu^2$ and/or $m$ by a coupling of the
fields to some set of interactions which dynamically generate a scale,
typically breaking a discrete $R$ symmetry, e.g.
\beq
{X W_\alpha^2 \over M} + X A^2 +{ S^2 \over M}AY
\eeq
The interactions associated with $W_\alpha^2$,$S$, generate a the
terms $\mu^2$ and $m$, and also an expectation value for $W$, of order
$FM$.  It would seem natural to identify $M$ with $M_p$, but in order
to destroy domain walls, it is necessary that $M$ (and $\langle
W_\alpha^2 \rangle$) be smaller than that.  It is the other
interactions, responsible for the disappearance of the domain walls,
which must generate the term which gives small cosmological constant.

\section{Gravity Waves From Domain Wall Collisions}
\label{gravitywaves}

Domain walls generally produce gravity waves when they collide and
disappear.  In this section we estimate the abundance and frequency of
the gravity waves, following Ref.~\cite{Takahashi:2008mu}.

In the violent collisions of domain walls, gravity waves are
produced at a frequency $f_p$ corresponding to a typical physical
length scale.  One of the important scales is the curvature radius of
the walls.  The domain-wall network is known to follow scaling
solution~\cite{PreRydSpe89,Hindmarsh:1996xv}, and then a typical
curvature radius is the Hubble horizon. Also, since the domain-wall
energy density decreases more slowly than radiation, most of the
gravity waves are produced when the walls disappear. Therefore we
expect that the gravity waves are produced at $f_p \sim H_d$, where
$H_d$ is the Hubble parameter at the disappearance of the
walls. Whether or not the gravity waves with frequencies greater than
$H_d$ are produced depends on the domain-wall dynamics at the
sub-horizon scales. In particular, there is another length scale, the
wall width, $\Delta$, which might affect the spectrum. According to
the numerical simulation~\cite{Hiramatsu:2010yz}, the gravity waves
from the walls have a broad and comparatively flat spectrum, ranging
from $f_p \sim H_d$ to $f_p \sim \Delta^{-1}$, with an intensity
consistent with that obtained from a simple dimensional analysis.

The frequency is red-shifted due to subsequent cosmic expansion, and
so, the frequency we observe today, $f_0$, is much smaller than that
at the production, $f_p$:
\beq
f_0
&\simeq&1\times10^2\, \lrfp{g_*}{200}{-\frac{1}{12}} \lrfp{H_d}{1\,{\rm GeV}}{-\frac{1}{2}} \lrf{f_p}{1\,{\rm GeV}}  {\rm Hz},
\label{freq}
\eeq
where  $g_*$ counts the relativistic degrees of freedom.
Here, radiation domination is assumed.

The intensity of the gravitational waves decreases as the universe
expands, since the amplitude too is red-shifted for sub-horizon modes.
In order to characterize the intensity, it is customary to use a
dimensionless quantity, $\Omega_{\rm gw}(f)$, defined by
\beq
\Omega_{\rm gw}(f) \;\equiv\; \frac{1}{\rho_c} \frac{d \rho_{gw}}{d \log f},
\eeq
where $\rho_{gw}$ is the energy density of the gravitational waves,
$\rho_c$ the critical energy density, and $f$ the frequency.  Let us
estimate the magnitude of $\Omega_{\rm gw}(f)$.  The energy of the
gravitational waves in a horizon at the disappearance of the walls is
estimated by
\beq
E_{gw} \;\sim\; G\, \frac{M_{DW}^2}{R_*} \sim  \frac{\sigma^2}{H_d^3 M_p^2},
\label{egw}
\eeq
where $G=1/(8 \pi M_p^2)$ is the Newton constant, $M_{DW}$ the energy
stored in the domain walls, and $R_*$ the typical spatial scale of the
energy distribution.  In the second equality in (\ref{egw}), we have
used $M_{DW} \sim \sigma / H_d^2$ and $R_* \sim 1/H_d$, where
$H_d \sim \epsilon/\sigma$ is the Hubble parameter when the walls disappear.
 If the reheating is completed before $H = m_{3/2}$,
we have
\beq
\Omega_{\rm gw}(f_0) h^2 \;\sim\; 10^{-5}
\lrfp{\sigma^2}{\epsilon M_p^2}{2},
\label{omega1}
\eeq
where $h$ is the present Hubble parameter in units of
$100$\,km/s/Mpc. Note that the condition for the domain walls to
disappear before the domination, (\ref{condition1}), is equivalent to
$\sigma^2/\epsilon M_p^2 \ll 1$.

So far, radiation domination has been assumed. The gravitino problem
implies however that the reheating temperature cannot be arbitrarily
high~\cite{Kawasaki:2004qu,Jedamzik:2006xz,de Gouvea:1997tn}.  If the
gravity waves are produced before the reheating,
$f_0$ is shifted to a smaller value and the intensity is
weakened. Specifically, assuming that the inflaton behaves like
non-relativisitic matter before the reheating, both the frequency and
intensity are suppressed by a factor of $(T_R^2/H_dM_p)^{3/2}$, where
$T_R$ is the reheating temperature.

Here let us briefly mention the sensitivities of the ongoing and
planned experiments on gravitational waves. One of the
ground-based experiments, LIGO~\cite{Abramovici:1992ah}, is in
operation and it is sensitive to the frequency between $O(10)$\,Hz and
$10^4$\,Hz. The latest upper bound is $\Omega_{\rm gw} h^2 < 6.5
\times 10^{-5}$ around $100$\,Hz~\cite{recent-LIGO}, and an upgrade of
the experiment, Advanced LIGO~\cite{Barish:1999vh,Fritschel:2003qw},
would reach sensitivities of $O(10^{-9})$. The sensitivity of
LCGT~\cite{Kuroda:2002bg} would be more or less similar to that of
Advanced LIGO.  There are also planned space-borne interferometers
such as LISA~\cite{LISA}, BBO~\cite{Crowder:2005nr} and
DECIGO~\cite{Seto:2001qf}. LISA is sensitive to the band of
$(0.03-0.1)\,{\rm mHz} \lsim f_0 \lsim 0.1$\,Hz, and it can reach
$\Omega_{\rm gw} h^2 < 10^{-12}$ at $f_0 = 1$\,mHz.  Moreover, BBO and
DECIGO will cover $10{\rm\,mHz} \lsim f_0 \lsim 10^2$\,Hz with much
better sensitivity.

Let us take the example considered in Sec.~\ref{annihilation}, i.e.,
the domain walls associated with the gaugino condensation at $\Lambda$
is destroyed due to a bias $\epsilon$ induced by a constant $w_c$~\cite{Takahashi:2008mu}.
In this case the walls disappear at $H_d \sim m_{3/2}$. Here we assume $b \ll a \sim 1$. The gravity waves have a broad spectrum from $f \sim
1\,{\rm kHz}\, (m_{3/2}/100{\rm GeV})^{1/2}$ to $f \sim 10^{14} \,{\rm
  Hz}\, b^{1/3} (m_{3/2}/100{\rm GeV})^{-1/6}$, with an intensity
$\Omega_{\rm gw}h^2 \sim 10^{-5} \,b^2$.  Therefore, for a light
gravitino mass $m_{3/2} \lsim 100$GeV and a moderately large value of
$b\, (\ll 1)$, the frequency and intensity may fall in the range
of the future gravity wave experiments.
In the second example, it is the domain walls associated with the
gaugino condensate, which gives a small cosmological constant,
disappear due to an explicit $R$-breaking induced by another gaugino condensate,
\beq
\int d^2 \theta \frac{S}{M} (W_\alpha^2+W_\alpha^\prime{}^2),
\eeq
where $S$ is a singlet, and $M \ll M_p$. In this case the tension and bias are given by $\sigma = \Lambda^3 \simeq
m_{3/2}M_p^2$ and $\epsilon \simeq \Lambda^3 \Lambda^{\prime 3}/M^2 \equiv c\, m_{3/2}^2 M_p^2$.
In order for the walls to disappear before they dominate the energy density of the Universe, $c$ must be greater than $1$.
The walls decay when $H_d \sim \Lambda^{\prime 3}/M^2 = c m_{3/2}$.   The gravity wave spectrum
ranges from $f \sim 1 \times 10^2\,{\rm Hz}\, c^{1/2} (m_{3/2}/{\rm
  GeV})^{1/2}$ to $f \sim 2\times10^{14} \,{\rm Hz}\, (m_{3/2}/{\rm
  GeV})^{-1/6}$, with an intensity $\Omega_{\rm gw}h^2 \sim 10^{-5}
\,c^{-2}$.

To summarize this section,  gravity waves are generally produced
when domain walls collide and disappear.  The frequencies of these gravity
waves happens to be close to those covered by the ongoing and planned
gravity-wave experiments.  Considering the sensitivities of future
experiments, the gravity waves from domain-wall decay may be
detectable, for certain parameters (e.g., not too small $b$ or
$c^{-1}$ in the cases considered above).

\section{Conclusions}
\label{conclusions}
There are a number of reasons to believe that discrete R symmetries
may play an important role in low energy supersymmetry. Perhaps most
dramatically, the smallness of $\la W \ra$ can be naturally explained
by a spontaneously broken discrete R symmetry, but such symmetries
are alsolikely to play a role in supersymmetry breaking, and may well be
important in understanding the suppression of rare process.  Such
discrete symmetry breaking implies the existence of domain walls.
Because of the role of $W$ as an order parameter for $R$ symmetry
breaking, and in accounting for the small value of the cosmological
constant, one has some idea of the tension of these walls.  We have
seen that adopting a picture for supersymmetry breaking then
determines whether these walls may or may not be produced before the
end of inflation.  In many scenarios, the domain walls can be inflated
away even by inflation at scales comparable to the GUT scale, but in
many others, they are produced after inflation.  In this case, they
must somehow disappear. This can be accomplished if there is a large
enough explicit $R$ breaking, and such breaking is plausible if one
examines typical string vacua with discrete $R$ symmetries.  In such
a case, there is typically a non-anomalous symmetry broken at a very
high scale, and an approximate symmetry broken at a lower scale.  It is the
domain walls associated with the lower scale breaking that are the greatest
danger.  But
requiring that the walls annihilate before gravitational collapse
provides strong constraints.    The interactions responsible for the
explicit $R$ breaking must, in particular, make the dominant
contribution to the superpotential.  In the case that high scale instantons
or analogous effects are responsible for the explicit breaking, these
must also give the dominant contribution to $W$; similarly, in
the case of two gaugino condensates, the couplings of the higher scale condensate must be suppressed
by a scale smaller than the Planck scale. If such domain walls were produced
at an early stage, we have seen that their annihilations have implication
for future gravity-wave experiments.

\noindent
{\bf Acknowledgements:}
FT thanks T.~Hiramatsu, M.~Kawasaki, and K.~Saikawa for discussion on the numerical simulation of
domain walls. The work of  FT was supported by the Grant-in-Aid for Scientific Research on Innovative Areas (No. 21111006)
and Scientific Research (A) (No. 22244030) and JSPS Grant-in-Aid for Young Scientists (B) (No. 21740160).
This work was supported by World Premier International Center Initiative (WPI Program), MEXT, Japan.
MD was supported in part by the U.S. Department of Energy, grant number DE-FG03-92ER40689.
He thanks Stanford University and the Stanford Institute for Theoretical Physics for a visiting faculty appointment while much of this work was performed.

\end{document}